# A reproducible 3D convolutional neural network with dual attention module (3D-DAM) for Alzheimer's disease classification

Gia Minh Hoang[1][0000-0002-8494-0096], Youngjoo Lee[1] and Jae Gwan Kim[1*][0000-0002-1010-7712]

[1] Department of Biomedical Science and Engineering, Gwangju Institute of Science and Technology, Republic of Korea.

**Abstract.** Alzheimer's disease is one of the most forms of dementia disease characterized by the accumulation of amyloid-beta plaque and tau tangles. Nowadays, Deep learning approaches have been widely used as promising techniques for Alzheimer's disease diagnosis. In this study, we propose a reproducible model using 3D convolutional neural networks with a dual attention module for Alzheimer's disease classification. We trained the model in the ADNI database and verified the generalizability of our method in two independent datasets (AIBL and OASIS1). Our method achieved state-of-the-art classification performance: 91.78% accuracy for MCI progression classification and 98.18% accuracy for Alzheimer's disease classification in the ADNI dataset. The generalizability performances are 86.37% accuracy in the AIBL dataset and 83.42% accuracy in the OASIS1 dataset. The experimental results show that the proposed approach has competitive performances in comparison with recent studies in terms of accuracy performance and generalizability. Using explainable AI, we also found that the hippocampus and temporal lobe were the strongest predictors of our model for Alzheimer's disease classification.

## 1 INTRODUCTION

Alzheimer's disease is one of the most common forms of neurodegenerative disease among old people that can progressively cause memory impairment, and greatly affect the activities of daily living. More than 6.7 million Americans aged 65 or older are living with Alzheimer's disease. By 2060, this number could rise to nearly 13.8 million [1]. In Alzheimer's disease patients, the brain starts to shrink, called progressive cerebral atrophy. Progression of atrophy is first manifested in the medial temporal lobe and then closely followed by the hippocampus, amygdala, and para-hippocampus [2]. There is no effective way to cure Alzheimer's disease, but there are treatments that may treat disease symptoms or delay disease progression. Therefore, the exact diagnosis of Alzheimer's disease may allow patients to start early treatment to slow or stop the progression of the disease.

With the development of neuroimaging techniques, brain atrophy could be used as one of the biomarkers for Alzheimer's disease diagnosis by visualizing MRI scans (best with a T1-weighted scan). The MRI scan could obtain the structure change in the brain with high resolution and contrast among white matter, gray matter, or small essential structures such as the hippocampus or amygdala. However, the diagnosis using MRI by the doctor is time-consuming and dependent on the expert's experience which could lead to a high misdiagnosing rate. Therefore, accurate computer-aid approaches for AD diagnosis are needed for clinical application. Recently, various studies have shown that deep learning models with MRI-based convolutional neural networks could classify Alzheimer's disease and Cognitive normal with considerable results [3], [4], [5], [6]. Although those studies achieved promising performance, there are still some limitations in their methodology. First, Alzheimer's disease and Cognitive Normal classification is a late diagnosis and cannot allow patients to start timely intervention. The progression of Mild cognitive impairment (MCI) to Alzheimer's disease is more important since MCI is the intermediate between Cognitive Normal and Alzheimer's disease. The MCI patients who do not have any treatment or medications could progress to Alzheimer's disease after 3 years. Second, there are lack of studies to prove the generalization of their approaches in independent datasets [7], [8], [9], [10]. A low generalization approach could have over-optimistic performance in particular datasets, but when applying those approaches to other datasets, the performance will dramatically fall. Therefore, the evaluation of the models in an independent dataset is crucial for clinical application.

The Convolutional Block Attention Module (CBAM) was first introduced by Woo et al. in 2018 at the European Conference on Computer Vision Conference [11]. The module included two main components: spatial attention and channel attention which can learn 'what' and 'where' to attend in the channel and spatial axes respectively. The CBAM has been widely applied in the medical field , especially medical imaging classification with promising results [12], [13], [14]. In our study, due to the 3D structure of whole brain MRI scans, we applied dual attention to our convolutional neural networks to capture the slice information across the channel and structural information across spatial axes.

In this study, we proposed the 3D convolutional neural networks with dual attention module (3D-DAM) for Alzheimer's disease classification by whole brain-based MRI scan. We make the following contributions. We apply a dual attention module to 3D-CNN architecture to classify Alzheimer's disease and achieve state-of-the-art performance in respect of accuracy in comparison with recent studies. We also achieved promising results when testing the generalizability of our model in the independent dataset. Using Explainable AI, we visualize the brain regions most affect to our methods. The hippocampus and temporal lobe were highlighted by attention score in our approaches, in agreement with previous studies.

3## 2 MATERIALS AND METHOD

### 2.1 DATASET DESCRIPTION AND IMAGE PREPROCESSING

In this study, the 1.5 T T1-weighted MRI images were collected from the public database of Alzheimer's Disease Neuroimaging Initiative (ADNI). In this study, we selected 403 Alzheimer's disease patients (AD), 653 Cognitive Normal patients (CN), 165 progressive Mild cognitive impairment patients (pMCI) and 205 stable Mild cognitive impairment patients (sMCI) from ADNI database. Table 1 shows the patient demographic of the studied subjects.

**Table 1.** Patient demographics of ADNI database.

|      | Subject | Session | Age           | Gender       |
|------|---------|---------|---------------|--------------|
| AD   | 403     | 1241    | 75.65 ± 7.8   | 225 M/ 178 F |
| CN   | 653     | 2564    | 76.15 ± 4.9   | 277 M/ 376 F |
| pMCI | 165     | 825     | 75.44 ± 7.27  | 104 M/ 64 F  |
| sMCI | 205     | 922     | 74.64 ± 5.14  | 121 M/ 84 F  |

The ADNI data have been curated and converted to the Brain Imaging Data Structure (BIDS) format following the processing pipeline of Clinica software [15] to avoid data leakage concerns pointed out by [16] such as wrong data split, late split., etc. The MRI scan selection pipeline has been shown in Fig. 1.A. The images then are pre-processed to obtain better image features for classification. In the ADNI database, image registration is needed to reduce global differences between each image among datasets. These registrations are following the t1-linear pipeline of Clinica [15]. First, bias field correction was applied using the N4ITK method [17]. Next, an affine registration was performed using the SyN algorithm [18] from ANTs [19] to align each image to the MNI space with the ICBM 2009c nonlinear symmetric template. Fig. 1.B shows an example of MRI scan before and after pre-processing.

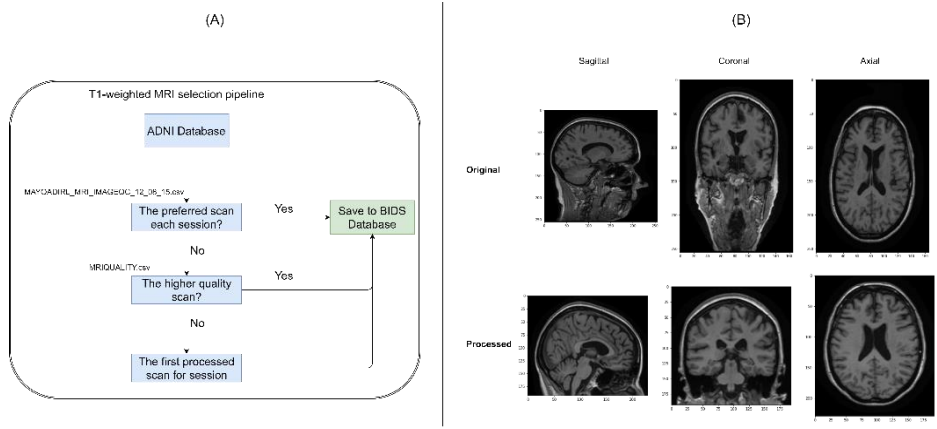

**Fig. 1.** MRI scan selection pipeline following adni-to-bids pipeline of Clinical. (A) The MRI scan selection pipeline. (B) Example of scan before and after preprocessing.

## 2.2   DUAL ATTENTION MODULE

Inspired by Convolutional Block Attention Module (CBAM) [11], we applied a dual attention module to our 3D-CNN model to capture the slice information across the channel and structural information across spatial axes. First, we produce a channel attention map by discovering the inter-channel relationship of features. As each feature map channel is considered a feature detector, channel attention focuses on the 'what' slice is meaningful given an input MRI scan. Given an intermediate feature map $F \in R^{C \times H \times W \times D}$ as input, we use average-pooling and max-pooling operations to generate two different spatial information of a feature map: $F_{avg}^c$ and $F_{max}^c$, respectively.

$$F_{avg}^c = AvgPool(F) \text{ and } F_{max}^c = MaxPool(F) \tag{1}$$

Then a shared multi-layer perceptron (MLP) network is applied to F1 and F2. After that, we merge the output feature vectors using element-wise summation. The channel attention is computed as:

$$F_{out}^c = \sigma(MLP(F_{avg}^c) + MLP(F_{max}^c)) \tag{2}$$
$$F' = F \otimes F_{out}^c \tag{3}$$

where σ denotes the sigmoid function, MLP is shared multi-layer perceptron, $F_{avg}^c$ and $F_{max}^c$ are average-pooling and max-pooling, respectively.

Secondly, the output from the channel attention module will be passed through the spatial attention module. Two different pooling along the channel axis (channel max pooling and channel average pooling) are applied to generate two feature maps: $F_{mean}^s$ and $F_{max}^s$ respectively. Then the two feature maps are concatenated as input for the subsequent convolutional layer. The output of the convolutional layer then passes through the sigmoid layer and could be considered a spatial attention map. In short, the spatial attention is computed as:

$$F_{mean}^s = torch.mean(F') \tag{4}$$

$$F_{max}^s = torch.max(F') \quad (5)$$

$$F_{out}^s = \sigma(Conv3D(F_{mean}^s + F_{max}^s)) \quad (6)$$

$$F'' = F' \otimes F_{out}^s \quad (7)$$

where σ denotes the sigmoid function, Conv3D is convolutional layer, $F'$ is output of channel attention module, $F_{avg}^c$ and $F_{max}^c$ are average-pooling and max-pooling, respectively. Fig. 2. describes the details of the Dual attention module (DAM).

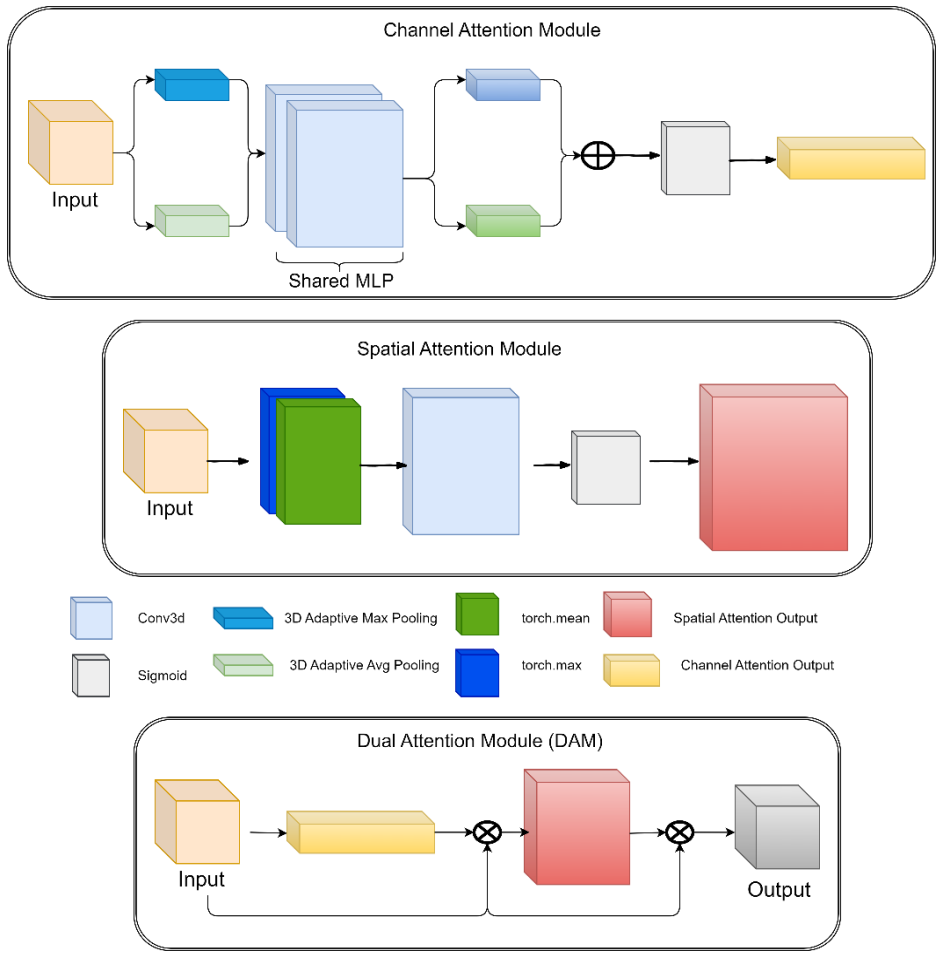

**Fig. 2.** Illustration of the Dual Attention Module (DAM).

## 2.3 3D CONVOLUTIONAL NEURAL NETWORKS WITH DUAL ATTENTION MODULE

Our proposed 3D convolutional neural network with a dual attention module (3D-DAM) framework is shown in Fig. 3. The backbone of the proposed network consists of 3D convolutional layers and dual attention module (DAM) layers. The first convolutional layer has a kernel size of 8×8×8. The next two layers are the residual block which included two convolutional layers with 16x16x16 and 32x32x32 kernel sizes, respectively. Each convolutional layer is followed by batch normalization (BN) and rectified linear unit (ReLU) activations. Then DAM layer was applied before features were passed through another residual block with a kernel size of 64x64x64. The other DAM module was applied then the features were pooled by the Avg Pooling layer. Finally, the Fully Connected layer has been used for classification.

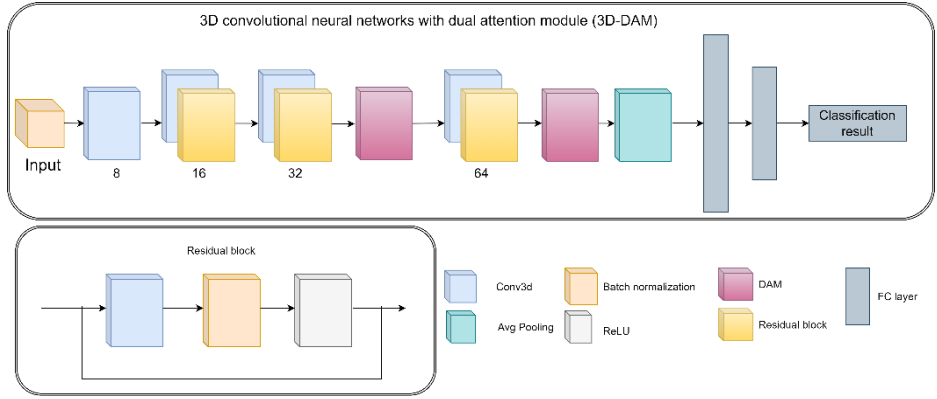

**Fig. 3.** Illustration of our 3D convolutional neural networks with dual attention module (3D-DAM).

## 2.4 EXPERIMENTAL SETTING AND EVALUATION METRICS

The proposed architecture is implemented using Python based on the PyTorch package, on a computer with an Intel(R) Xeon(R) Gold 6258R CPU @ 2.70GHz with 256GB RAM. The GPU used is 4x NVIDIA GeForce RTX 3090. We trained the network using the SGD optimizer with a first momentum of 0.9 and the second momentum of 0.999. The initial learning rate and L2-Regularization value are set to $10^{-7}$ and $10^{-6}$, respectively. We set the maximum number of training epochs to 200 and used a batch size of 16 at each iteration.

We have evaluated two classification tasks of AD classification (AD vs. NC) and MCI conversion prediction (pMCI vs. sMCI). We employ the classification accuracy (Acc), sensitivity (Sen), specificity (Spec), to evaluate the performance of the classification model, calculated by the following formulas:

$$Acc = \frac{TP+TN}{TP+TN+FP+FN} \qquad (7)$$

$$Sen = \frac{TP}{TP+FN} \tag{8}$$

$$Spec = \frac{TN}{TN+FP} \tag{9}$$

where TP is True positive, TN is True negative, FP is False positive, FN is False Negative.

## 3 RESULTS

### 3.1 CLASSIFICATION PERFORMANCE ON ADNI

The performances on AD classification and MCI progression classification achieved by our 3DDAM method and the other methods on the test set from ADNI are shown in Table 2 and Fig. 4. As shown in Table 2, our method achieved 98.18% of accuracy, 96.73% of sensitivity, and 98.97% of specificity in AD classification. In terms of MCI progression classification, we achieved 91.78% accuracy, 95.06% sensitivity, and 88.62% specificity. Our proposed method achieved a top-ranked classification performance in both AD classification and MCI conversion tasks in most cases.

**Table 2.** AD Classification (AD vs. CN) and MCI Conversion Prediction (pMCI vs. sMCI) performances on the ADNI Test Set.

| Reference | AD vs CN | | | pMCI vs sMCI | | |
|---|---|---|---|---|---|---|
| | ACC | SEN | SPEC | ACC | SEN | SPEC |
| [20] | 97.35 | 97.10 | 97.95 | 78.79 | 75.16 | 82.42 |
| [21] | 92.4 | 91.0 | 93.8 | 80.2 | 77.1 | 82.6 |
| [22] | 97.78 | 95.59 | 99.82 | 79.90 | 75.55 | 99.70 |
| [23] | - | - | - | 87.8 | 88 | 88 |
| [6] | 91.3 | 91.0 | 91.9 | 82.1 | 81.2 | 80.9 |
| Our Method | 98.18±0.42 | 96.73±1.08 | 98.97±0.51 | 91.78±0.63 | 95.06±1.11 | 88.62±1.15 |

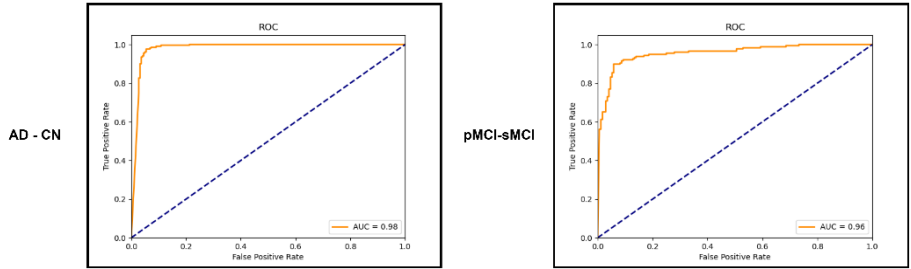

**Fig. 4.** Classification result for the proposed model for AD vs CN and pMCI vs sMCI classification.

### 3.2 THE GENERALIZABILITY EVALUATION USING AIBL AND OASIS

To investigate the generalizability of our proposed model, we have conducted an evaluation on two independent datasets (AIBL and OASIS1). The patient demographics of the AIBL and OASIS1 datasets were shown in Table 3.

**Table 3.** Patient demographics of the AIBL and OASIS1 database.

| Data  | Type | Subject | Age           | Gender       |
|-------|------|---------|---------------|--------------|
| AIBL  | AD   | 77      | 84 ± 25.7     | 34 M/ 43 F   |
|       | CN   | 449     | 73.96 ± 6.5   | 187 M/ 262 F |
| OASIS | AD   | 100     | 76.76 ± 7.11  | 41 M/ 59 F   |
|       | CN   | 316     | 65 ± 13.89    | 119 M/ 197 F |

In this evaluation, we only performed AD vs CN classification task due to insufficient pMCI and sMCI samples obtained from AIBL and OASIS1 datasets. We perform two evaluation experiments as follow:

Experiment 1: We trained our proposed model based on the ADNI dataset and evaluated it using two independent datasets.

Experiment 2: We reversed the training and test datasets in which we use AIBL and OASIS1 as training and ADNI as the test set.

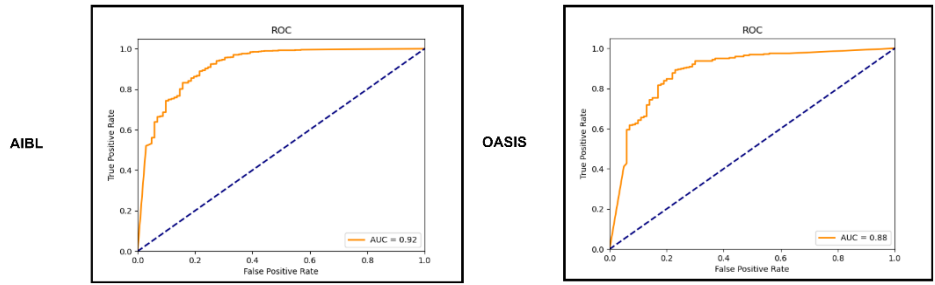

**Fig. 5.** Classification result for the AIBL and OASIS dataset using ADNI as training set.

The classification performance of experiment 1 has been shown in Table 4, and Figure 5. The 3D-DAM method obtains results with potential on all three metrics (i.e., Acc = 86.3%, Sen = 80.2%, and Spec = 87.1%) in AD classification in the AIBL dataset. The classification results in OASIS dataset are 83.4%, 85.8%, 82.6% in accuracy, sensitivity, specificity respectively using the model trained on the ADNI dataset. In Experiment 2, we trained our model with AIBL and OASIS dataset and evaluated in ADNI set. The classification results are 85.4%, 80.1% and 89.5% in Accuracy, Sensitivity, Specificity respectively.

**Table 4.** Generalizability evaluation result of Experiment 1 and Experiment 2.

| Train | Evaluation | Acc | Sen | Spec |
| --- | --- | --- | --- | --- |
| ADNI | AIBL | 86.3 | 80.2 | 87.1 |
| ADNI | OASIS | 83.4 | 85.8 | 82.6 |
| AIBL-OASIS | ADNI | 85.4 | 80.1 | 89.5 |

### 3.3 PATHOLOGICAL BRAIN REGION BY 3DDAM MODEL

One of the most important tasks for computer-aid diagnosis is defining the pathological region that the model focuses on. In this study, we use the GradCam to investigate which brain region is the most crucial part of our 3DDAM model to classify AD and CN. Fig. 6 depicts the sagittal, coronal, and axial plane slices with attention score overlay to demonstrate the highest attention region. As shown in Fig. 6, the hippocampus, medial temporal lobe, and amygdala are three brain regions that play important roles in classifying AD and CN of the proposed model. These regions are consistent with many previous studies on AD diagnosis [2], [24], [25].

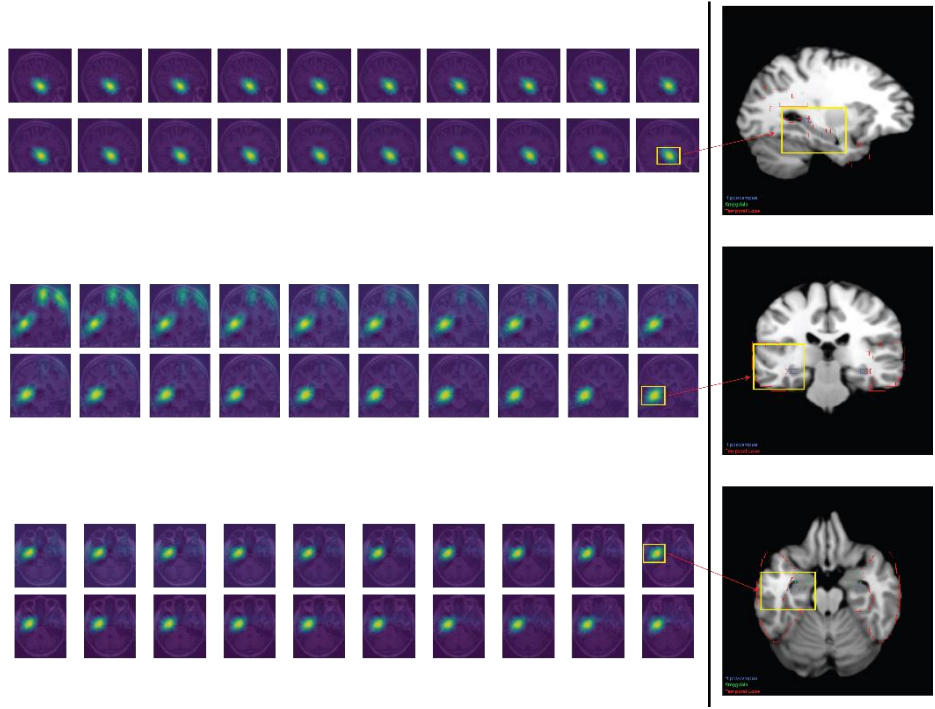

**Fig. 6.** Visualization of the pathological brain region identified by the proposed method of AD Classification. The left panel shows the informative locations suggested by attention scores. The right panel shows the related brain region, respectively, with marked locations.

## 4   DISCUSSION

Computer-aided approaches to Alzheimer's disease are crucial for early intervention as Alzheimer's disease pathology is irreversible and has no effective treatment. the MRI scan could acquire high spatial resolution and contrast brain atrophy as a pathology sign for Alzheimer's disease diagnosis. In this study, we proposed a dual attention-based deep learning framework to classify the progression of MCI to AD. Our approaches achieved state-of-the-art performance compared with recent MRI-based deep learning studies [21], [22], [23]. These results suggested that channel and spatial attention could improve the traditional convolutional neural networks in terms of Alzheimer's disease classification. In our opinion, the dual attention mechanism could effectively capture the disease-related structural changing features of the brain in both channel and spatial dimensions.

Application of the Deep Learning model in medical clinical is still challenging due to its difficulty in reproducibility and generalizability. To test our model generalizability, we use two independent datasets to evaluate our method. We have promising performance when evaluating AIBL and OASIS1 using the ADNI-trained model and vice versa. The evaluation result indicates the good generalization capability of our method for AD diagnosis. The Acc, Sen, and Spec of our proposed method based on ADNI are slightly higher than other datasets. The main reason is the ADNI dataset and the AIBL as well as the OASIS1 dataset are collected from distinct protocols, which could have different signal-to-noise ratios and device parameters.

Although our proposed method achieves good performance in AD classification and identifying pathological brain regions, there are still limitations that may affect the reproducible capability of our model. First, convolution layers to extract feature may be difficult to capture different levels of details from source images, especially in MCI progression classification. In MCI patients, the brain atrophy between stable and progressed MCI is not significantly different; therefore, the CNN power may be still not powerful enough to capture all these changes. Secondly, for clinical use, MRI scans from different hospitals may have different settings and sizes of images. In our proposed method, the size of the input scan is fixed. The fixed size model could not be widely used in clinical. Thirdly, while the reproducibility and generalizability of AD classification have been proven by AIBL and OASIS1 dataset, MCI progression classification needs more data to verify its generalizability. In future work, we can embed more novel networks instead of basic 3D CNN to MRI scan feature extraction. The combination of MRI features and clinical data could be a promising solution to improve our performance in the future. For generalizability improvement, other national cohorts should be applied to verify its capacity to use in clinical.

## 5 CONCLUSION

Inspired by the dual attention mechanism and 3D convolutional neural networks, we propose a 3D convolutional neural network with a dual attention module (3D-DAM) approach to advance computer-aid Alzheimer's disease diagnosis, which includes three major contributions. Our proposed method evaluated the largest database of Alzheimer's disease (ADNI) and verified the generalizability on other independent datasets (AIBL and OASIS1). We achieved an Alzheimer's disease classification accuracy of 98.18%, sensitivity of 96.73%, and specificity of 98.97%. In terms of MCI progression classification, we achieved 91.78% accuracy, 95.06% sensitivity, and 88.62% specificity. Our proposed method also showed good generalizability performance in AIBL and OASIS1 for Alzheimer's disease classification tasks. The explainable AI can identify and highlight the highest attention brain region for model decisions such as the hippocampus and medial temporal lobe. Future work will focus on the continuous improvement of model performance and generalizability using more independent datasets.

# References

1. "2023 Alzheimer's disease facts and figures," Alzheimer's & Dementia, vol. 19, no. 4, pp. 1598–1695, 2023, doi: 10.1002/alz.13016.
2. K. A. Johnson, N. C. Fox, R. A. Sperling, and W. E. Klunk, "Brain Imaging in Alzheimer Disease," Cold Spring Harb Perspect Med, vol. 2, no. 4, p. a006213, Apr. 2012, doi: 10.1101/cshperspect.a006213.
3. G. Mukhtar, "Convolutional Neural Network Based Prediction of Conversion from Mild Cognitive Impairment to Alzheimer's Disease: A Technique using Hippocampus Extracted from MRI," Adv. Electr. Comp. Eng., vol. 20, no. 2, Art. no. 2, May 2020, doi: 10.4316/AECE.2020.02013.
4. S. Ahmed, B. C. Kim, K. H. Lee, H. Y. Jung, and for the A. D. N. Initiative, "Ensemble of ROI-based convolutional neural network classifiers for staging the Alzheimer disease spectrum from magnetic resonance imaging," PLOS ONE, vol. 15, no. 12, p. e0242712, Dec. 2020, doi: 10.1371/journal.pone.0242712.
5. H. Li, M. Habes, D. A. Wolk, Y. Fan, and Alzheimer's Disease Neuroimaging Initiative and the Australian Imaging Biomarkers and Lifestyle Study of Aging, "A deep learning model for early prediction of Alzheimer's disease dementia based on hippocampal magnetic resonance imaging data," Alzheimer's & Dementia, vol. 15, no. 8, pp. 1059–1070, 2019, doi: 10.1016/j.jalz.2019.02.007.
6. X. Zhang, L. Han, W. Zhu, L. Sun, and D. Zhang, "An Explainable 3D Residual Self-Attention Deep Neural Network for Joint Atrophy Localization and Alzheimer's Disease Diagnosis Using Structural MRI," IEEE Journal of Biomedical and Health Informatics, vol. 26, no. 11, pp. 5289–5297, Nov. 2022, doi: 10.1109/JBHI.2021.3066832.
7. X. Feng, F. A. Provenzano, S. A. Small, and for the Alzheimer's Disease Neuroimaging Initiative, "A deep learning MRI approach outperforms other biomarkers of prodromal Alzheimer's disease," Alzheimer's Research & Therapy, vol. 14, no. 1, p. 45, Mar. 2022, doi: 10.1186/s13195-022-00985-x.
8. S. Fathi, M. Ahmadi, and A. Dehnad, "Early diagnosis of Alzheimer's disease based on deep learning: A systematic review," Computers in Biology and Medicine, vol. 146, p. 105634, Jul. 2022, doi: 10.1016/j.compbiomed.2022.105634.
9. A. Loddo, S. Buttau, and C. Di Ruberto, "Deep learning based pipelines for Alzheimer's disease diagnosis: A comparative study and a novel deep-ensemble method," Computers in Biology and Medicine, vol. 141, p. 105032, Feb. 2022, doi: 10.1016/j.compbiomed.2021.105032.
10. G. M. Hoang, U.-H. Kim, and J. G. Kim, "Vision transformers for the prediction of mild cognitive impairment to Alzheimer's disease progression using mid-sagittal sMRI," Frontiers in Aging Neuroscience, vol. 15, 2023, Accessed: Feb. 19, 2024. [Online]. Available: https://www.frontiersin.org/articles/10.3389/fnagi.2023.1102869
11. S. Woo, J. Park, J.-Y. Lee, and I. S. Kweon, "CBAM: Convolutional Block Attention Module." arXiv, Jul. 18, 2018. doi: 10.48550/arXiv.1807.06521.
12. Z. Zhang and M. Wang, "Convolutional Neural Network with Convolutional Block Attention Module for Finger Vein Recognition." arXiv, Feb. 14, 2022. doi: 10.48550/arXiv.2202.06673.
13. Y. Zhang, X. Zhang, and W. Zhu, "ANC: Attention Network for COVID-19 Explainable Diagnosis Based on Convolutional Block Attention Module," CMES, vol. 127, no. 3, pp. 1037–1058, 2021, doi: 10.32604/cmes.2021.015807.
14. S. Wang et al., "Improved Multi-Stream Convolutional Block Attention Module for sEMG-Based Gesture Recognition," Frontiers in Bioengineering and Biotechnology, vol. 10, 2022,

Accessed: Feb. 19, 2024. [Online]. Available: https://www.frontiersin.org/articles/10.3389/fbioe.2022.909023
15. A. Routier et al., "Clinica: An Open-Source Software Platform for Reproducible Clinical Neuroscience Studies," Frontiers in Neuroinformatics, vol. 15, 2021, Accessed: Feb. 19, 2024. [Online]. Available: https://www.frontiersin.org/articles/10.3389/fninf.2021.689675
16. J. Wen et al., "Convolutional neural networks for classification of Alzheimer's disease: Overview and reproducible evaluation," Medical Image Analysis, vol. 63, p. 101694, Jul. 2020, doi: 10.1016/j.media.2020.101694.
17. N. J. Tustison et al., "N4ITK: Improved N3 Bias Correction," IEEE Transactions on Medical Imaging, vol. 29, no. 6, pp. 1310–1320, Jun. 2010, doi: 10.1109/TMI.2010.2046908.
18. B. B. Avants, C. L. Epstein, M. Grossman, and J. C. Gee, "Symmetric diffeomorphic image registration with cross-correlation: Evaluating automated labeling of elderly and neurodegenerative brain," Medical Image Analysis, vol. 12, no. 1, pp. 26–41, Feb. 2008, doi: 10.1016/j.media.2007.06.004.
19. B. B. Avants, N. J. Tustison, M. Stauffer, G. Song, B. Wu, and J. C. Gee, "The Insight ToolKit image registration framework," Frontiers in Neuroinformatics, vol. 8, 2014, Accessed: Feb. 19, 2024. [Online]. Available: https://www.frontiersin.org/articles/10.3389/fninf.2014.00044
20. J. Zhang, B. Zheng, A. Gao, X. Feng, D. Liang, and X. Long, "A 3D densely connected convolution neural network with connection-wise attention mechanism for Alzheimer's disease classification," Magnetic Resonance Imaging, vol. 78, pp. 119–126, May 2021, doi: 10.1016/j.mri.2021.02.001.
21. W. Zhu, L. Sun, J. Huang, L. Han, and D. Zhang, "Dual Attention Multi-Instance Deep Learning for Alzheimer's Disease Diagnosis With Structural MRI," IEEE Transactions on Medical Imaging, vol. 40, no. 9, pp. 2354–2366, Sep. 2021, doi: 10.1109/TMI.2021.3077079.
22. M. Ashtari-Majlan, A. Seifi, and M. M. Dehshibi, "A Multi-Stream Convolutional Neural Network for Classification of Progressive MCI in Alzheimer's Disease Using Structural MRI Images," IEEE Journal of Biomedical and Health Informatics, vol. 26, no. 8, pp. 3918–3926, Aug. 2022, doi: 10.1109/JBHI.2022.3155705.
23. C. Wang et al., "A high-generalizability machine learning framework for predicting the progression of Alzheimer's disease using limited data," npj Digit. Med., vol. 5, no. 1, Art. no. 1, Apr. 2022, doi: 10.1038/s41746-022-00577-x.
24. E. J. Burton et al., "Medial temporal lobe atrophy on MRI differentiates Alzheimer's disease from dementia with Lewy bodies and vascular cognitive impairment: a prospective study with pathological verification of diagnosis," Brain, vol. 132, no. 1, pp. 195–203, Jan. 2009, doi: 10.1093/brain/awn298.
25. N. Schuff et al., "MRI of hippocampal volume loss in early Alzheimer's disease in relation to ApoE genotype and biomarkers," Brain, vol. 132, no. 4, pp. 1067–1077, Apr. 2009, doi: 10.1093/brain/awp007.